# Disease X vaccine production and supply chains: risk assessing healthcare systems operating with artificial intelligence and industry 4.0

Petar Radanliev[1] · David De Roure[1]

**Abstract**
**Objective** The objective of this theoretical paper is to identify conceptual solutions for securing, predicting, and improving vaccine production and supply chains.
**Method** The case study, action research, and review method is used with secondary data – publicly available open access data.
**Results** A set of six algorithmic solutions is presented for resolving vaccine production and supply chain bottlenecks. A different set of algorithmic solutions is presented for forecasting risks during a Disease X event. A new conceptual framework is designed to integrate the emerging solutions in vaccine production and supply chains. The framework is constructed to improve the state-of-the-art by intersecting the previously isolated disciplines of edge computing; cyber-risk analytics; healthcare systems, and AI algorithms.
**Conclusion** For healthcare systems to cope better during a disease X event than during Covid-19, we need multiple highly specific AI algorithms, targeted for solving specific problems. The proposed framework would reduce production and supply chain risk and complexity in a Disease X event.

**Keywords** Disease X · Vaccine production and supply chains · Risk assessment · Healthcare systems · Artificial intelligence · Industry 4.0

## 1 Introduction

Covid-19 has triggered a chain of technological developments in healthcare systems that lead to the creation of 'Society 5.0' [1], where artificial intelligence (AI) and industry 4.0 present the basis for digital personalised healthcare. Digitalising the healthcare system has proven beneficial for managing Covid-19 but has also increased the cyber risk surface, while the benefits and improvements for vaccine production and supply remain to be seen. This article conceptualises the construction of predictive algorithms [2] that can improve the cyber risk posture of health care systems and the ability to rapidly adapt to huge external changes. Because *'We are likely on the edge of a major transformation in how many of us live – and how goods are produced and distributed.'* [3]. The conceptual approach enables undertaking experimental developments on progressing from the current state of risk assessment into measuring risk with analytical methods. These experimental research developments are focused on the: (1) need to classify Covid-19 risk data into primary and secondary data sets for training predictive algorithms; (2) the need to include AI in healthcare networks that intersect a diverse set of isolated production and supply chain domains. The main motivating point for this article is the timing - global pandemics such as Covid-19 are very rare events. Although there are several viruses in circulation (e.g., Zika, AIDS), it has been a century since the last major global pandemic - the Spanish flu. Thus, we can say that investigating Covid-19 is once in a lifetime event and the findings will be of utmost value in dealing with Disease X.

✉ Petar Radanliev
  petar.radanliev@oerc.ox.ac.uk

1 Oxford e-Research Centre, Department of Engineering Sciences, University of Oxford, Oxford, UK

## 1.1 Review of vaccine production and supply chain bottlenecks

We have learned valuable lessons from analysing past pandemics and viruses such as Influenza, yet the production and supply process for Influenza vaccines still takes months. To prepare the healthcare system for future pandemics, we need to improve the production and supply chain capacity and speed. One of the commonly discussed solutions is to stockpile medications, but that is not happening. For example, we have existing antivirals to combat viruses like Tamiflu for Influenza, but there are no sufficient stockpiles of antivirals in the world for even one country. Without stockpiling, the current production and supply chains will face many difficulties and bottlenecks in a Disease X scenario.

To prevent vaccine production and supply chain bottlenecks, this section conducts a conceptual analysis of six potential bottlenecks and suggests six solutions that encompass a design process for constructing AI algorithms based on new and emerging data and technologies during a Disease X event. First solution $S_1$: since the supply chain will be delivering a high-value, highly-needed vaccine, firstly the design process needs to concentrate on securing the deliveries from cyber-risk, theft, sabotage, counterfeiting, etc. The first solution is targeted at the first potential bottleneck, the risk of deliberate tampering. Second solution $S_2$: is targeted for the second bottleneck and incorporates revision of the personal shortages and the requirement to vaccinate and protect from dangers the individuals operating the entire supply/value chain, before starting a large-scale vaccination. Triggering questions on who will deliver a vaccine to the people delivering the vaccines, and how long will that take. Thirst solution $S_3$: is targeted at investigating the lack of coordination and risk from shortages caused by the lack of infrastructure to monitor the vaccination records. $S_4$: Fourth is to identify a potential shortage of supplies in critical materials during Disease X, and these can be something meaningless, like rubber stoppers for the vials. $S_5$: Fifth is to construct algorithmic optimisations and solutions for the supply chain bottlenecks that cause limited capacity, e.g., lack of cold chain storage and equipment failure, resulting in vaccine damage. This is even more concerning for rural areas and small towns, with small hospitals. $S_6$: The final bottleneck is to categorise different misinformation strategies and fake news during Disease X, (e.g., microchips in the vaccines) and prepare counter strategies. In some societies, almost 60% of the population is not planning to take the vaccine, and we need over 70% vaccination for a pandemic to subdue. Fake news spread six times faster than scientific news. This bottleneck places a great deal of focus on the dissemination of results and targeting different audiences with real-life stories and narratives.

In the first stage, the production and supply chains should focus on addressing these bottlenecks. In the second stage, the AI system should enable the development of quantitative risk, impact and value assessment with real-time data, allowing for dynamic optimisation and managing of a Disease X event.

## 1.2 Integrating algorithms in cyber risk assessment of healthcare systems

The integration of algorithms in cyber risk impact assessment of failures in healthcare systems - in the same way that was used to forecast Covid-19 growth rates [5], would enable a progression from manual to automated assessment. The integration of AI in risk assessment presents a novel concept for advancing from physical to digital healthcare systems and progressing from qualitative to quantitative assessment. Such algorithms can be designed with a quantitative 'causal-comparative design' for synthesising data on actions and events that occurred during Covid-19. This solution can predict the risks and enhance our preparedness for Disease X.

A quantitative *'correlational research'* can be applied to assess the statistical relationship between the *primary* and *secondary* failures in healthcare systems, and to measure the actual loss [6]. This requires classifying probabilistic data into *primary* - Disease X events and the health system response/failure to the events, and the *secondary* - the failure of other health systems as the reactions to Disease X event (e.g., delayed surgeries, delayed cancer diagnosis, and/or treatment). This classification will be a significant improvement on the current state-of-the-art in cyber loss assessment which is based on qualitative methodologies. At present, there are no methods, frameworks, models, or formulas for measuring the secondary risk from cyber events, not to mention extraordinary events that could trigger catastrophic loss of life - such as Disease X. But as we have seen from the Covid-19, the secondary loss is not something that we can keep ignoring in our risk assessments. For example, if a cyber event disrupts the vaccine cold chain, resulting in a loss of a few thousand vaccines, then the primary risk is the loss of the vaccines, and that can easily be calculated and measured. The secondary risk is a different story, we need to consider the loss of life not only from Covid-19, but also from many other cascading events e.g., the medical staff being occupied for longer with vaccinations, and unable to perform other crucial healthcare work, resulting with more loss of life. Since there are no mechanisms for reporting cyber-attacks, and the associated damage from the event, finding probabilistic data is a challenge, especially the *secondary* risk data. This is one of the reasons why similar research has not been done in the past.

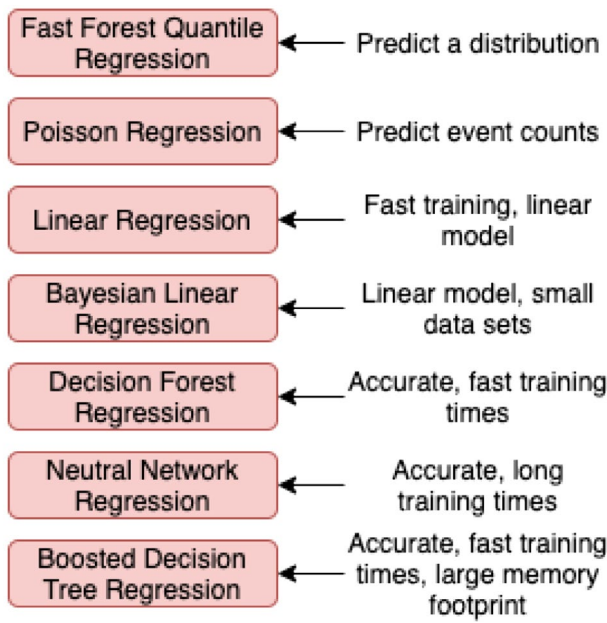

**Fig. 1** Regression – algorithmic solution for predicting and forecasting risks from a Disease X event

### 1.3 New forms of data for assessing risk and optimising the vaccine production and supply chains

The separation of *primary* and *secondary* risk is developed upon past experiences in terms of the (un)availability of data. This concern can be addressed by expanding new and emerging forms of data (e.g., linking digital behavioural data (e.g., social media) with survey data). The integration of algorithms in risk assessing and optimising the vaccine production and supply chain can apply *stratified sampling* and *random sampling* and integrate the FAIR Institute method[1] for the data analysis and the comparative aspects of the collected data. The difference between disciplines means that we need to adapt the model with a focus on healthcare systems and not on financial systems, which is the primary goal of the FAIR method. This adaptation can be conducted simultaneously with the advancement of the model with AI algorithms for automated risk assessment (Fig. 1). The automation approach should include unsupervised learning algorithms, which would enable testing the model randomly, at different stages, and in different settings. Secondly, a qualitative *case study* and *red-teaming* approach should be used, to challenge and test the state-of-the-art in healthcare cybersecurity. Although failure and compromise will occur, the objective of applying red teaming would be to find solutions to how the system responds in these circumstances.

[1] https://www.fairinstitute.org/about.

**Table 1** Six solutions for vaccine production and supply chain bottlenecks

| M | Algorithmic methods (M) emerging as solutions (S) to production and supply chain bottlenecks in Disease X event | S |
|---|---|---|
| $M_1$ | Securing the vaccine supply chain for Disease X: design algorithms for quantitative risk analytics with new and emerging forms of data. | $S_1$ |
| $M_2$ | Constructing alternative vaccine delivery systems based on new technologies e.g., drones, autonomous vehicles, and robots. | $S_2$ |
| $M_3$ | Algorithmic design for dynamic coordination and self-adapting predictive (real-time) analytics of cyber risks in Disease X supply chains. | $S_3$ |
| $M_4$ | Identifying how modern technologies can help with resolving shortages of supplies in critical times e.g., 3D printing [4]. | $S_4$ |
| $M_5$ | Constructing supply chain models based on modern technologies real-time data and adaptive algorithms. | $S_5$ |
| $M_6$ | Validation of the security readiness for Disease X: designing a self-adapting AI system that can continue operating when compromised by a cyber-attack. | $S_6$ |

The six solutions in Table 1 are designed for securing digital health systems operating at the edge of the network. The algorithmic solutions in Fig. 1 are synthesised for predicting cyber risk magnitude through dynamic analytics of cyber-attack threat event frequencies. The conceptual framework integrating the vaccine production and supply chain solutions in Fig. 2 is constructed to improve the state-of-the-art by intersecting the previously isolated disciplines, such as confidence intervals; time-bound ranges. The solution in Fig. 2 is constructed by applying a qualitative case study - action research on the knowledge we accumulated of Covid-19. The solutions interconnect contemporary mindsets on data privacy and security [7]. These are all novel concepts and approaches that lead to development between disciplines and expand beyond the current state of the art. To apply the new solutions, we need to develop a series of algorithms and build upon each algorithm to reach a state of AI that can perform to the satisfaction of our healthcare systems.

## 2 Discussion

The expected difficulties in applying the new solutions include the lack of data for testing the algorithms on a real-time vaccine supply chain – healthcare providers might be reluctant to allow experimental technology in their operational supply chains. Although we identified numerous testbeds (e.g., UK digital catapults), all of the testbeds are based in controlled environments, and Disease X can be unpredictable. Even if I could secure a real-world testbed, the pandemic might be finished before the algorithms are operational, disabling the ability to test and verify the outcomes in real-world/real-time scenarios. In addition, expected difficulties include the modification of the healthcare systems' physical parameters as a result of Disease X, for example, during Covid-19 many new hospitals were built at speed and the cybersecurity of these field hospitals was not a priority, given the serious and urgent risk of Covid-19 on the loss of life. Assessing with precision the adaptation mechanisms without an understanding of the exact system setup will be difficult. The future researchers might get some insights from analysing Covid-19 on how to resolve this. Alternatives to mitigate this risk include (1) seeking existing real-world testbeds designed for similar purposes – by healthcare providers, governments, private organisations; research institutions; (2) building a small-scale low-cost supply chain testbed for this project - with a limited number of connected devices – mostly sourced from existing testbeds built for different purposes. The second solution would be quite relevant for developing countries (e.g., Africa) where low-cost technology is often used in the medical supply chains. In addition, the second solution enables the project to test the algorithms more extensively. But the algorithms will not be tested in a practical scenario, resulting in a lack of understanding of their performance in an unexpected situation.

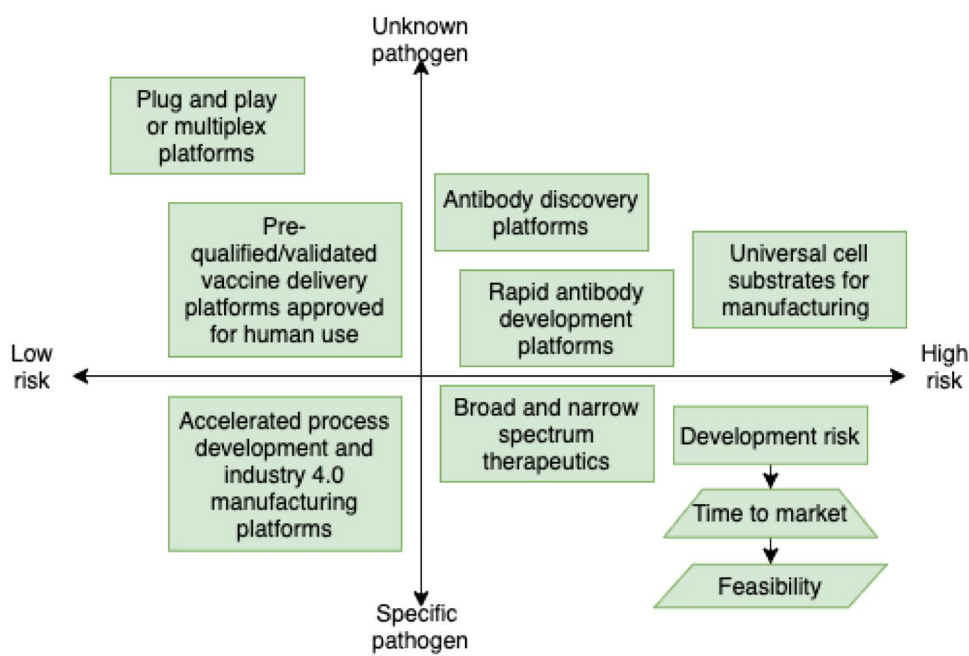

**Fig. 2** Conceptual framework integrating the new solutions for production and supply chain response during a Disease X event

## 3 Conclusion

Covid-19 has caused a significant advancement in digital health, triggered by the advantages of remote management of the pandemic. The desperate need for fast and remote healthcare data sharing attracted an increased investment in internet-connected health systems and devices, but cybersecurity and the cyber-risk of these complex and coupled systems have not been investigated with a great deal of scrutiny.

This article engaged with undertaking tentative and experimental conceptual developments on examining how algorithms can secure, optimise, predict (forecast) failures and improve the vaccine production and supply chains in Disease X crises. Without appropriate risk assessment of the increasingly more digitalised healthcare systems, the potential risks could exceed the benefits, especially because of the expected difficulties in integrating legacy healthcare systems with complex AI networks.

Until present, we have trained AI algorithms to learn what we already know, and in managing Covid-19 we are trying to train algorithms (e.g., with unsupervised learning) to predict what we don't know – namely, insights on Covid-19 and future Disease X events. To mitigate the risk of creating a single point of failure, we need to design multiple highly specific AI algorithms, targeted for solving specific problems, then connect the output. This approach would present lower risk and less complexity.

What we have learned from Covid-19 is even with our advanced healthcare systems, global pandemics can spread undetected until it is too late to be contained. This article presents a state-of-the-art solution based on edge analytics, for detecting anomalies and predicting (i.e., forecasting) a Disease X event, and the associated risks to the healthcare system, including cyber risks.

### 3.1 Limitations and further research

The expected difficulties in applying the new solutions include addressing multiple objectives at speed - automated AI depends on data preparation and data might be difficult to obtain. In addition, there could be an incompatibility between the new and emerging forms of data and the AI algorithm training requirements. Alternatives for future researchers to mitigate this risk include using training data from Covid-19 and editing the parameters to reflex the characteristics of Disease X. The new AI solutions are based on experimental developments in research on AI algorithms for risk assessment from adversaries that attack the healthcare system during Disease X. Future experimental research should be focused on the (1) need to train the new cybersecurity AI algorithm to stop AI-driven cyber-attacks; (2) need to train AI algorithm how to prevent active and passive reconnaissance by adversarial AI - about a specific target at scale.

Additional expected difficulties on a testing and validation level include the need to test the performance of the algorithm in measuring risks and observe the effectiveness in the presence and absence of a deadly and fast-spreading virus. This could be difficult because of the unpredictable duration of Covid-19 the pandemic can be finished before the new algorithms are developed, disabling the ability to test the algorithms in real-time. Alternatives to mitigate this risk include testing the algorithms with existing data from Covid-19 and verifying their effectiveness as solutions for the next pandemic and Disease X. This will mean that the algorithms developed during Covid-19, would be tested on training data limited by the parameters of (by then) past pandemics, and will need readjusting of the training data to include an element of randomness - which is expected from Disease X.

**Acknowledgements** Eternal gratitude to the Fulbright Visiting Scholar Project.

**Author contribution** All authors contributed equally.

**Funding** This work was funded by the EPSRC [grant number: EP/S035362/1] and by the Cisco Research Centre [grant number CG1525381].

**Data availability** all data and materials included in the article.

**Code availability** N/A – no code was developed; code was however used for running the R Studio analysis.

### Declarations

**Competing interests** On behalf of all authors, the corresponding author states that there is no conflict nor competing interest.